\documentclass[runningheads]{llncs}
\usepackage[T1]{fontenc}
\usepackage[utf8]{inputenc}
\usepackage{rotating}
\usepackage[final]{microtype}
\usepackage{color}
\usepackage{latexsym}
\usepackage{amssymb}
\usepackage{amsmath}
\usepackage{mathtools}
\usepackage{xspace}
\usepackage{multirow}
\usepackage{relsize}
\usepackage{wrapfig}
\usepackage{bussproofs}
\usepackage{graphicx}
\usepackage{lipsum}
\usepackage[hypertexnames=false]{hyperref}
\usepackage[caption=false, position=top]{subfig}
\usepackage{turnstile}
\usepackage{centernot}
\usepackage{placeins}

\usepackage{algorithm2e}
\RestyleAlgo{boxed}
\LinesNumbered
\SetKwInput{KwInput}{Input}
\SetKwInput{KwOutput}{Output}
\SetKwRepeat{Do}{do}{while}

\newcommand{\qrattool}{\textsf{QRATPre\texttt{+}}\xspace}

\newcommand{\eabs}{\mathit{Abs}\xspace}
\newcommand{\unitprop}{\mathrel{\vdash\kern-.65em_{^{_{1}}}}\xspace}
\newcommand{\unitQprop}{\mathrel{\vdash\kern-.75em_{^{_{1\forall}}}}}
\newcommand{\nunitQprop}{\mathrel{\nvdash\kern-.75em_{^{_{1\forall}}}}}

\newcommand{\outerres}{\mathsf{OR}\xspace}

\newcommand{\qrat}{\mathsf{QRAT}\xspace}

\newcommand{\qat}{\mathsf{QAT}\xspace}
\newcommand{\newqrat}{\mathsf{QRAT}^{+}\xspace}
\newcommand{\qbce}{\mathsf{QBCE}\xspace}
\newcommand{\ble}{\mathsf{BLE}\xspace}

\newcommand{\textnewqrate}{\mathsf{QRATE}^{+}\xspace}

\newcommand{\textnewqratu}{\mathsf{QRATU}^{+}\xspace}

\newcommand{\rn}{\mathsf{RN}\xspace}

\newcommand{\prefix}{\Pi}
\newcommand{\clauset}{\psi}
\newcommand{\qclauset}{\phi}

\newcommand{\hqspre}{\textsf{HQSpre}\xspace}
\newcommand{\qute}{\textsf{Qute}\xspace}
\newcommand{\bloqqer}{\textsf{Bloqqer}\xspace}
\newcommand{\rareqs}{\textsf{RAReQS}\xspace}
\newcommand{\caqe}{\textsf{CAQE}\xspace}
\newcommand{\caqefixed}{\textsf{CAQE}\xspace}
\newcommand{\depqbf}{\textsf{DepQBF}\xspace}

\newcommand{\ijtihad}{\textsf{Ijtihad}\xspace}

\begin{document}

\newcommand{\doctitle}{\qrattool: Effective QBF Preprocessing via Strong Redundancy Properties}

\title{ \doctitle\thanks{Part of this work was carried out while the first author was employed
  at the Institute of Logic and Computation, TU Wien, Austria. This work is supported by the Austrian Science Fund (FWF) under grant
  S11409-N23 and will appear in the \textbf{proceedings} of SAT~2019, LNCS, Springer, 2019.}}

\author{Florian Lonsing\inst{1} \and
Uwe Egly\inst{2} 
}

\authorrunning{}

\institute{Computer Science Department, Stanford University, Stanford, CA 94305, USA
  \\
  \and
  Institute of Logic and Computation, TU Wien, 1040 Vienna, Austria \\
}

\maketitle

\begin{abstract}
We present version 2.0 of \qrattool, a preprocessor for quantified
Boolean formulas (QBFs) based on the $\qrat$ proof system and its
generalization $\newqrat$.  These systems rely on strong redundancy properties of clauses and
universal literals.  \qrattool is the first implementation of these
redundancy properties in $\qrat$ and $\newqrat$ used to simplify QBFs in 
preprocessing. It is written in C and features an API for easy
integration in other QBF tools. We present implementation details and
report on experimental results demonstrating that \qrattool improves upon the power of
state-of-the-art preprocessors and solvers.
\end{abstract}


\section{Introduction}

The application of preprocessing prior to the actual solving process
is crucial for the performance of most of the quantified Boolean
formula (QBF)
solvers~\cite{DBLP:conf/cp/LonsingE18,Lonsing201692,DBLP:journals/fuin/MarinNPTG16}. Preprocessors
aim at decreasing the complexity of a given formula with respect to
the number of variables, the number of clauses, or the number of
quantifier blocks. Contrary to complete QBF solvers, preprocessors are
incomplete but can detect redundant parts of the given formula
by applying resource-restricted reasoning. 
\bloqqer~\cite{DBLP:journals/jair/HeuleJLSB15} and
\hqspre~\cite{DBLP:conf/tacas/WimmerRM017} are leading preprocessors
which show their power in the yearly QBFEVAL competitions\footnote{\url{http://www.qbflib.org/index_eval.php}} in 
potentially almost doubling the number of instances solved by certain
state-of-the-art
solvers. These tools 
apply a diverse set of redundancy elimination techniques.

We present \qrattool~2.0,\footnote{\qrattool is licensed
  under GPLv3: \url{https://lonsing.github.io/qratpreplus/}} 
a QBF preprocessor based on the $\newqrat$ proof
system~\cite{DBLP:conf/cade/LonsingE18}. 
\qrattool processes QBFs in prenex conjunctive normal form (PCNF) and
eliminates redundant clauses and universal literals.  Redundancy
checking relies on redundancy properties defined by the $\newqrat$
proof system, which is a generalization of the $\qrat$ (quantified
resolution asymmetric tautology) proof
system~\cite{DBLP:conf/cade/HeuleSB14,DBLP:journals/jar/HeuleSB17}. $\qrat$
is a lifting of
(D)RAT~\cite{DBLP:conf/cade/JarvisaloHB12,DBLP:conf/sat/WetzlerHH14}
from the propositional to the QBF level, and it simulates virtually
all simplification rules applied in state-of-the-art QBF preprocessors. 
This is made possible by strong redundancy \nolinebreak properties.

However, the strong redundancy properties of $\qrat$ have not been
applied to preprocess QBFs so far. With \qrattool we close this gap
and leverage the power of $\qrat$ and $\newqrat$ for QBF
preprocessing.  Compared to the initially released version 1.0,
version 2.0 comes with a more modularized code base and a C API that
allows to easily integrate \qrattool in other tools. \qrattool currently applies
only rewrite rules of the $\qrat$ and $\newqrat$ proof systems that
remove either clauses or universal literals from a PCNF. That is, it
does not attempt to add redundant parts with the aim to potentially enable further
simplifications later on. Despite this fact, experimental results with
benchmarks from QBFEVAL'18 clearly indicate the effectiveness of
\qrattool. It improves on state-of-the-art preprocessors, such as \bloqqer and \hqspre, and solvers in terms of
formula size reduction and solved instances, respectively.


\section{$\newqrat$ Redundancy Checking for Preprocessing}
\label{sec:qrat:workflow}

\qrattool eliminates redundant clauses and
universal literals within clauses from a QBF in
PCNF. Redundancy checking in \qrattool relies on redundancy
properties defined by the $\newqrat$ proof system. We
present $\newqrat$ only informally and refer to related  
work instead~\cite{DBLP:conf/cade/LonsingE18}.

Let $\qclauset := \prefix. (\clauset \wedge (C' \cup \{l\}))$ be a PCNF
with \emph{prefix} $\prefix := Q_1B_1 \ldots Q_iB_i \ldots
\linebreak \ldots Q_nB_n$, where $Q_iB_i$ are \emph{quantifier blocks
  (qblocks)} consisting of a quantifier $Q_i \in \{\forall,
\exists\}$ and a block (i.e., set) $B_i$ of variables.
We write $Q(B_1 \ldots  B_n)$ for $Q(B_1 \cup \ldots \cup B_n)$.
Index $i$ is
the \emph{nesting level} of qblock $Q_iB_i$ and of the variables in $B_i$. Formula $\clauset \wedge
(C' \cup \{l\})$ is in CNF, where $(C' \cup \{l\})$ is a clause
in $\qclauset$ containing literal $l$.
We consider only PCNFs without tautological clauses of the form
$(C \cup \{p\} \cup \{\bar p\})$ for some propositional variable $p$. 

\begin{figure}[t]
\centering\framebox{\includegraphics[scale=0.9]{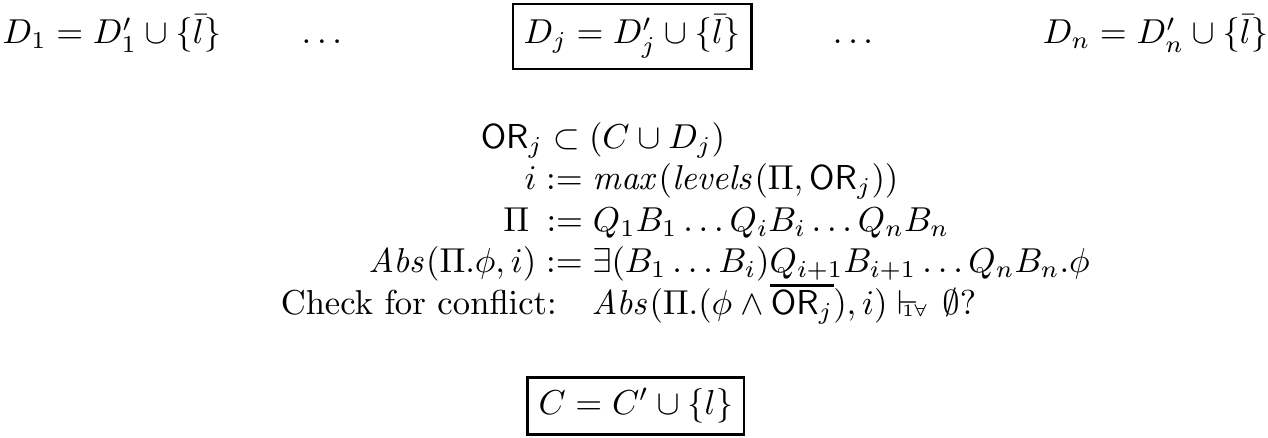}}
\caption{Redundancy checking of clause $C := (C' \cup \{l\})$ based on
  $\newqrat$. The resolution neighborhood $\rn(C,l)$ of $C$ consists
  of the clauses $D_i$ with $\bar l \in D_i$ shown on top. The boxes
  indicate the pair of clauses for which the current outer resolvent
  $\outerres_j$ is computed. }
\label{fig:qratplus}
\end{figure}

Given a clause $C := (C' \cup \{l\})$ in $\qclauset$, checking whether
$C$ or the literal $l \in C$ is redundant requires to consider all
clauses $D_i$ in the \emph{resolution neighborhood} $\rn(C,l) := \{D_i
\mid D_i \in \phi, \bar l \in D_i\}$ of $C$ with respect to $l$, cf.~\cite{DBLP:conf/cade/JarvisaloHB12,DBLP:conf/cade/KieslSTB16,DBLP:conf/ijcai/KieslSTB17}. This
is illustrated in Fig.~\ref{fig:qratplus}. 
Given $C$ and
some $D_j \in \rn(C,l)$, we first compute the \emph{outer resolvent}
$\outerres_j \subset (C \cup D_j)$ of $C$ and $D_j$ on literal $l$~\cite{DBLP:journals/jar/HeuleSB17}. Then we
determine the maximum nesting level $i :=
\mathit{max}(\mathit{levels}(\prefix, \outerres_j))$ of variables
appearing in $\outerres_j$. Based on $i$ we construct the PCNF
$\eabs(\prefix. \clauset, i) := \exists (B_1 \ldots B_i) Q_{i+1}B_{i+1} \ldots Q_n
B_n. \clauset$ by converting all universal quantifiers in the subprefix 
$Q_1B_1 \ldots Q_iB_i$ to existential ones. The resulting PCNF
$\eabs(\prefix. \clauset, i)$ is an \emph{abstraction} of
$\prefix. \clauset$.

We add the negation $\overline{\outerres_j}$ of the outer resolvent
$\outerres_j$, which is a set of unit clauses, to the abstraction $\eabs(\prefix. \clauset, i)$ to
obtain the formula $\eabs(\prefix. (\clauset \wedge
\overline{\outerres_j}), i)$. Finally, we check whether a conflict,
i.e., the empty clause $\emptyset$, is derived by applying \emph{QBF
  unit
  propagation (QBCP)}~\cite{DBLP:conf/aaai/CadoliGS98,DBLP:journals/jair/GiunchigliaNT06,DBLP:conf/tableaux/Letz02,DBLP:conf/iccad/ZhangM02}
to $\eabs(\prefix. (\clauset \wedge \overline{\outerres_j}), i)$. If so, which we denote by $\eabs(\prefix. (\clauset \wedge \overline{\outerres_j}),
i) \unitQprop \emptyset$, then the current outer resolvent
$\outerres_j$ has the \emph{quantified asymmetric tautology
  (QAT)}~\cite{DBLP:conf/cade/LonsingE18} redundancy property.

If \emph{all possible} outer resolvents of $C := (C' \cup \{l\})$ and
clauses $D$ in the resolution neighborhood $\rn(C,l)$ of $C$ with
respect to literal $l$ have the $\qat$ property, then clause $C$ has the \emph{$\newqrat$ redundancy property} on literal $l$. In this case, either $C$ or
$l$ is redundant, depending on whether $l$ is existential or
universal, respectively. Note that in general, for every outer
resolvent $\outerres_j$, the index $i$ for which the abstraction
$\eabs(\prefix. (\clauset \wedge \overline{\outerres_j}), i)$ is
constructed may be different.

Eliminating redundant clauses or universal literals via the above workflow is denoted by $\textnewqrate$ and
$\textnewqratu$, respectively. In \qrattool, we apply the $\textnewqrate$ and
$\textnewqratu$ rewrite rules for preprocessing. 
The $\qrat$ redundancy property~\cite{DBLP:journals/jar/HeuleSB17}
differs from $\newqrat$~\cite{DBLP:conf/cade/LonsingE18} in that for
$\qrat$ always a full abstraction $\eabs(\prefix. (\clauset \wedge
\overline{\outerres_j}), i)$ with $i := n$ is constructed, 
regardless of
the actual maximum nesting level of variables in the current outer
resolvent $\outerres_j$. Moreover, $\newqrat$ relies on QBCP 
which includes the universal reduction
operation~\cite{DBLP:journals/iandc/BuningKF95} to temporarily shorten
clauses during propagation. This way, potentially more
conflicts are derived. In contrast to that, $\qrat$ applies propositional unit
propagation to abstractions where all variables are existential.  Due to that, the
$\newqrat$ redundancy property is more general and stronger than
$\qrat$.

\section{Implementation Details}
Algorithm~\ref{algo} shows the high-level workflow implemented in
\qrattool. To limit the computational costs of the relative expensive
techniques $\textnewqrate$ and $\textnewqratu$, which are based on QBCP, 
we apply cheaper ones first to reduce the formula
size upfront. In \emph{quantified blocked clause elimination
  (QBCE)}~\cite{DBLP:conf/cade/BiereLS11}, which is a restriction of
$\textnewqrate$, it is checked whether every outer resolvent
(cf.~Fig.~\ref{fig:qratplus}) contains a pair of complementary
literals. \emph{Blocked literal elimination
  (BLE)}~\cite{DBLP:conf/nfm/HeuleSB15} is a restriction of
$\textnewqratu$ and, like $\qbce$, checks for complementary literals
in outer resolvents. Before checking whether some clause $C$ has the
$\newqrat$ property, we check whether it has the $\qat$
property. This is done analogously to checking whether an outer
resolvent has the $\qat$ property in $\newqrat$ testing~(cf.~Section~\ref{sec:qrat:workflow}).
$\qat$ checking is necessary in the workflow since it is not subsumed by $\textnewqrate$.

During clause elimination, all
clauses found redundant in the current PCNF~$\qclauset'$ are
cleaned up lazily in one pass after an application of a
 technique. In 
\begin{wrapfigure}{r}{0.49\textwidth}
\begin{algorithm}[H]
  \label{algo}
  \SetAlgoLined
  \KwInput{\hspace*{0.275cm}PCNF $\qclauset$.}
  \KwOutput{Simplified PCNF $\qclauset'$.}
  
  $\qclauset' = \qclauset$\;
  \Do{$\qclauset'$ changed}
     {
       \tcp{clause elimination}
    $\qclauset' := \qbce(\qclauset')$\;  
    $\qclauset' := \qat(\qclauset')$\;
    $\qclauset' := \textnewqrate(\qclauset')$\; 
    \tcp{literal elimination}
    $\qclauset' := \ble(\qclauset')$\; 
    $\qclauset' := \textnewqratu(\qclauset')$\;
  }
  \caption{\mbox{\qrattool workflow.}}
  \end{algorithm}
  \vspace*{-0.80cm}
  \end{wrapfigure}
literal elimination, redundant literals are cleaned up
 eagerly since clauses are shortened, which increases chances to
 detect conflicts in QBCP.

If the $\newqrat$ redundancy check fails for some clause $C$
with literal $l\in C$ and clause $D_i \in \rn(C,l)$ in the resolution
neighborhood of $C$, we mark $D_i$ as a \emph{witness} for that
failure. If a witness $D$ is found redundant then all clauses in
$\rn(D,l)$, for all literals $l \in D$, are scheduled for being
checked in the next iteration as $D$ has prevented at least one of these
clauses from being detected before. This witness-based scheduling 
potentially avoids superfluous redundancy checks.  In our
experiments the median number of clauses being checked in a run of
\qrattool on a given PCNF was only by a factor of~3.3 larger than the
initial number of clauses in the PCNF.

We maintain the index $i$ indicating the maximum nesting level of
variables in the current outer resolvent being tested
(cf.~Fig.~\ref{fig:qratplus}). Any variable with an index smaller
than $i$ is treated as an existential one during QBCP. This way,
abstractions $\eabs(\prefix. \clauset, i)$ are constructed
implicitly. For QBCP, we implemented standard two-literal
watching~\cite{DBLP:conf/sat/GentGNRT03}. However, when assignments
are retracted after deriving a conflict, then in general the literal watchers have to be
restored to literals which are existential in the input PCNF rather
than in the current abstraction. Restoring literal watchers in our implementation is necessary to
maintain certain invariants, in contrast to, e.g., QBCP in QCDCL solvers.

Compared to version 1.0, version 2.0 of \qrattool comes with a C
API that allows for easy integration in other tools. The API provides
functions to import and export PCNFs and to configure the
preprocessing workflow. Options include switches to toggle the
individual redundancy tests in the main loop, user-defined limits, and
shuffling the orderings in which clauses are tested. Shuffling may
affect the result of preprocessing since the $\qat$, $\textnewqrate$ and
$\textnewqratu$ rewrite rules are not confluent. In the default
configuration, \qrattool does not shuffle clauses and applies rewrite
rules until saturation.


\section{Experiments}

\qrattool improves on state-of-the-art preprocessors and solvers in
terms of formula size reduction and solved
instances. We ran experiments with 
the preprocessors \bloqqer~v37~\cite{DBLP:journals/jair/HeuleJLSB15} and
\hqspre~1.3~\cite{DBLP:conf/tacas/WimmerRM017}, and the solvers
\caqe (commit-ID~9b95754 on GitHub)~\cite{DBLP:conf/fmcad/RabeT15,DBLP:conf/cav/Tentrup17}, 
\rareqs~1.1~\cite{Janota20161},
\ijtihad~v2~\cite{DBLP:conf/fmcad/BloemBHELS18},
\qute~1.1~\cite{DBLP:conf/sat/PeitlSS17}, and
\depqbf~6.03~\cite{DBLP:conf/cade/LonsingE17}. The solvers 
implement different solving paradigms, e.g., QCDCL
(\qute and \depqbf), expansion (\rareqs and \ijtihad), and clausal abstraction~(\caqe).

The following experiments were run on a cluster of Intel Xeon CPUs
(E5-2650v4, 2.20 GHz) running Ubuntu 16.04.1. We used the 463
instances from the PCNF track of QBFEVAL’18.  In all
our experiments, we allowed 600s CPU time and 7~GB of memory for each call of \bloqqer,
\hqspre, or \qrattool. For formulas where \bloqqer or \hqspre exceeded
these limits, we considered the original, unpreprocessed formula. In
contrast to \bloqqer and \hqspre, we implemented a soft time limit in
\qrattool which, when exceeding the limit, allows to print the
preprocessed formula with identified redundant parts being \nolinebreak removed.

Table~\ref{fig:exp:463:sizes} shows the effect of preprocessing
by \qrattool (Q), \bloqqer (B), \hqspre (H), and combinations, 
where \qrattool is called before (QB, QH) or after 
\begin{wraptable}{r}{0.51\textwidth}
\vspace*{-1.0cm}
\caption{Effect of preprocessing.}
\vspace*{0.1cm}
\label{fig:exp:463:sizes}
{\setlength\tabcolsep{0.15cm}
\begin{tabular}{lccccccc}
\hline
  & Q  & B  & H  & QB & BQ & QH & HQ\\
\hline
\emph{\#cl}         &  79 & 78 & 23 & 70 & 69 &  18 & 17  \\
\emph{\#qb}         &  92 & 22 & 17 & 21 & 22 &  59 & 17  \\
\emph{\#$\exists$l} &  82 & 85 & 27 & 80 & 77 &  20 & 21  \\
\emph{\#$\forall$l} &  73 & 95 & 74 & 82 & 86 &  44 & 53  \\
\hline
\end{tabular}
}
\vspace*{-0.6cm}
\end{wraptable}
(BQ, HQ) \bloqqer or \hqspre. The table shows 
\emph{average numbers} of clauses (\emph{\#cl}), qblocks
(\emph{\#qb}), existential (\emph{\#$\exists$l}) and universal
literals (\emph{\#$\forall$l}) as a percentage relative to the
original benchmark set. Except for qblocks, \qrattool
considerably further reduces the formula size when combined with both
\bloqqer (B vs.~QB and BQ) and \hqspre (H vs.~QH and
HQ). These results include solved instances: 
\qrattool, \bloqqer, and \hqspre solve 18, 74, and 158 original instances,
respectively. The preprocessors timed out on 12~(\hqspre),
38~(\bloqqer), and 59 (\qrattool, soft time limit) original \nolinebreak instances.

The application of computationally inexpensive techniques like $\qbce$
and $\ble$ to shrink the PCNF before applying more expensive ones like
$\textnewqrate$ and $\textnewqratu$ pays off. With the
original workflow (Algorithm~\ref{algo} and column Q in
Table~\ref{fig:exp:463:sizes}), \qrattool spends 96s on average per 
instance, compared to 120s when disabling $\qbce$ and $\ble$, where it
exceeds the time limit on 77 instances.

Shuffling the ordering of clauses before applying 
$\textnewqrate$ and $\textnewqratu$ in Algorithm~\ref{algo} based on
five different random seeds hardly had any effect on the aggregate
data in column Q in Table~\ref{fig:exp:463:sizes}, except for an
increase in eliminated universal literals by one percent.  When using
the redundancy property of $\qrat$, which is weaker than the one of
$\newqrat$, by constructing full abstractions
(cf.~Section~\ref{sec:qrat:workflow}), we observed a moderate decrease
of all four metrics by one percent \nolinebreak each.

\begin{table}[t]
\caption{QBFEVAL'18: solved instances (\emph{S}), unsatisfiable (\emph{$\bot$}), 
satisfiable (\emph{$\top$}), and uniquely solved ones (\emph{U}), and total CPU time in kiloseconds (K) including time outs.}
\addtocounter{table}{-1}
\begin{minipage}[b]{0.499\textwidth}
\begin{center}
\subfloat[Original instances.]{
{\setlength\tabcolsep{0.15cm}
\begin{tabular}{l@{\quad}r@{\quad}r@{\quad}r@{\quad}r@{\quad}c}
\hline
\emph{Solver} & \multicolumn{1}{l}{\emph{S}} & \multicolumn{1}{l}{\emph{$\bot$}} & \emph{$\top$} & \emph{U} & \emph{Time} \\
\hline
\caqefixed & 151 & 107 & 44 & 11 & 586K \\
\depqbf & 149 & 87 & 62 & 50 & 592K \\
\rareqs & 147 & 115 & 32 & 4 & 588K \\
\ijtihad & 132 & 111 & 21 & 2 & 609K \\
\qute & 98 & 79 & 19 & 6 & 665K \\
\hline
\end{tabular}
}
\label{fig:exp:463:instances:noprepro}
}
\end{center}
\end{minipage}
\hfill
\begin{minipage}[b]{0.499\textwidth}
\begin{center}
\subfloat[\qrattool only (Q).]{
{\setlength\tabcolsep{0.15cm}
\begin{tabular}{l@{\quad}c@{\quad}r@{\quad}c@{\quad}r@{\quad}r}
\hline
\emph{Solver} & \emph{S} & \multicolumn{1}{l}{\emph{$\bot$}} & \emph{$\top$} & \emph{U} & \emph{Time} \\
\hline
\caqefixed & 211 & 135 & 76 & 29 & 487K \\
\rareqs & 178 & 120 & 58 & 9 & 533K \\
\depqbf & 165 & 83 & 82 & 33 & 562K \\
\ijtihad & 156 & 111 & 45 & 1 & 562K \\
\qute & 137 & 92 & 45 & 9 & 598K \\
\hline
\end{tabular}
}
\label{fig:exp:463:instances:prepro}
}
\end{center}
\end{minipage}

\bigskip

\begin{minipage}[b]{0.499\textwidth}
\begin{center}
\subfloat[\bloqqer only (B).]{
{\setlength\tabcolsep{0.15cm}
\begin{tabular}{l@{\quad}r@{\quad}r@{\quad}r@{\quad}r@{\quad}r}
\hline
\emph{Solver} & \multicolumn{1}{l}{\emph{S}} & \multicolumn{1}{l}{\emph{$\bot$}} & \multicolumn{1}{l}{\emph{$\top$}} & \emph{U} & \emph{Time} \\
\hline
\caqefixed & 269 & 150 & 119 & 24 & 383K \\
\rareqs & 258 & 159 & 99 & 7 & 399K \\
\ijtihad & 200 & 128 & 72 & 4 & 482K \\
\depqbf & 198 & 99 & 99 & 21 & 501K \\
\qute & 189 & 114 & 75 & 2 & 512K \\
\hline
\end{tabular}
}
\label{fig:exp:463:instances:prepro:bloqqer}
}
\end{center}
\end{minipage}
\hfill
\begin{minipage}[b]{0.499\textwidth}
\begin{center}
\subfloat[\qrattool and \bloqqer (QB).]{
{\setlength\tabcolsep{0.15cm}
\begin{tabular}{l@{\quad}c@{\quad}c@{\quad}r@{\quad}r@{\quad}r}
\hline
\emph{Solver} & \emph{S} & \emph{$\bot$} & \multicolumn{1}{l}{\emph{$\top$}} & \emph{U} & \emph{Time} \\
\hline
\caqefixed & 290 & 169 & 121 &37  & 347K \\
\rareqs & 260 & 163 & 97 & 5 & 390K \\
\ijtihad & 216 & 140 & 76 & 2 & 456K \\
\depqbf & 210 & 107 & 103 & 21 & 481K \\
\qute & 197 & 119 & 78 & 2 & 493K \\
\hline
\end{tabular}
}
\label{fig:exp:463:instances:prepro:qrat:bloqqer}
}
\end{center}
\end{minipage}

\bigskip

\begin{minipage}[b]{0.499\textwidth}
\begin{center}
\subfloat[\hqspre only (H).]{
{\setlength\tabcolsep{0.15cm}
\begin{tabular}{l@{\quad}c@{\quad}c@{\quad}r@{\quad}r@{\quad}c}
\hline
\emph{Solver} & \emph{S} & \emph{$\bot$} & \multicolumn{1}{l}{\emph{$\top$}} & \emph{U} & \emph{Time} \\
\hline
\caqefixed & 322 & 183 & 139 & 21 & 290K \\
\rareqs & 292 & 180 & 112 & 2 & 317K \\
\depqbf & 270 & 160 & 110 & 20 & 364K \\
\qute & 253 & 161 & 92 & 3 & 390K \\
\ijtihad & 249 & 167 & 82 & 0 & 394K \\
\hline
\end{tabular}
}
\label{fig:exp:463:instances:prepro:hqspre}
}
\end{center}
\end{minipage}
\hfill
\begin{minipage}[b]{0.499\textwidth}
\begin{center}
\subfloat[\hqspre and \qrattool (HQ).]{
{\setlength\tabcolsep{0.15cm}
\begin{tabular}{l@{\quad}c@{\quad}c@{\quad}r@{\quad}r@{\quad}r}
\hline
\emph{Solver} & \emph{S} & \emph{$\bot$} & \multicolumn{1}{l}{\emph{$\top$}} & \emph{U} & \emph{Time} \\
\hline
\caqefixed & 325 & 188 & 137 & 14 & 279K \\
\rareqs & 303 & 189 & 114 & 3 & 304K \\
\depqbf & 271 & 158 & 113 & 20 & 362K \\
\qute & 263 & 170 & 93 & 2 & 377K \\
\ijtihad & 245 & 166 & 79 & 0 & 407K \\
\hline
\end{tabular}
}
\label{fig:exp:463:instances:prepro:hqspre:qrat}
}
\end{center}
\end{minipage}

\label{fig:exp:463:instances}
\refstepcounter{table}
\end{table}

Tables~\ref{fig:exp:463:instances:noprepro}
to~\ref{fig:exp:463:instances:prepro:hqspre:qrat} show the numbers of
instances solved after preprocessing with different combinations of
\qrattool, \bloqqer, and \hqspre. We used a limit of 1800s CPU time 
and 7~GB of memory for solving. Times for preprocessing are not included in the times
reported in the tables.  Preprocessing by \qrattool increases the
number of solved instances in most cases, except for \ijtihad on instances preprocessed with
\hqspre and \qrattool
(Tables~\ref{fig:exp:463:instances:prepro:hqspre}
and~\ref{fig:exp:463:instances:prepro:hqspre:qrat}).
Similar to formula size reduction shown in Table~\ref{fig:exp:463:sizes}, the
ordering of whether to apply \qrattool before or after \bloqqer or
\hqspre has an impact on 
solving performance. 
Interestingly,
for all solvers the combination BQ (\bloqqer before \qrattool) results in a decrease
of solved instances compared to QB (\qrattool before \bloqqer). We
made similar observations for combinations with \hqspre (QH and HQ).\footnote{We refer to the appendix for results with combinations BQ and QH.}


\section{Conclusion}

We presented version 2.0 of \qrattool, a preprocessor for QBFs in PCNF
that is based on strong redundancy properties of clauses and universal
literals defined by the $\newqrat$ proof
system~\cite{DBLP:conf/cade/LonsingE18}. $\newqrat$ is a
generalization of the $\qrat$ proof
system~\cite{DBLP:conf/cade/HeuleSB14,DBLP:journals/jar/HeuleSB17}. \qrattool
is the first implementation of the $\qrat$ and $\newqrat$ redundancy
properties for applications in QBF preprocessing.
As such, the techniques implemented in \qrattool are orthogonal to
techniques applied in state-of-the-art preprocessors like \bloqqer and \hqspre.
Our experiments 
demonstrate a considerable performance increase of
preprocessing and solving. \qrattool comes with a C
API that allows easy integration into other tools.

We observed a sensitivity of solvers to the ordering in which
\qrattool is coupled with other preprocessors. 
To better understand
the interplay between redundancy elimination and the
proof systems implemented in solvers, we want to further analyze this phenomenon. We used a simple but effective
witness-based scheduling to avoid superfluous redundancy checks. However, with more
sophisticated watched data structures the run time of \qrattool could
be optimized. To enhance the power of 
redundancy checking, it could be beneficial to
selectively add redundant formula parts to enable additional
simplifications \nolinebreak afterwards.



\newpage

\begin{appendix}

\section{Appendix}


\subsection{Additional Experimental Data}

\begin{table}[ht]
\caption{QBFEVAL'18: solved instances (\emph{S}), unsatisfiable (\emph{$\bot$}), 
satisfiable (\emph{$\top$}), and uniquely solved ones (\emph{U}), and total CPU time in kiloseconds (K) including time outs.}
\addtocounter{table}{-1}
\begin{minipage}[b]{0.499\textwidth}
  \begin{center}
\subfloat[\bloqqer and \qrattool (BQ).]{
{\setlength\tabcolsep{0.15cm}
\begin{tabular}{l@{\quad}c@{\quad}c@{\quad}r@{\quad}r@{\quad}r}
\hline
\emph{Solver} & \emph{S} & \emph{$\bot$} & \multicolumn{1}{l}{\emph{$\top$}} & \emph{U} & \emph{Time} \\
\hline
\caqefixed & 284 & 165 & 119 & 35 & 358K \\
\rareqs & 257 & 161 & 96 & 5 & 399K \\
\depqbf & 204 & 102 & 102 & 17 & 490K \\
\ijtihad & 197 & 127 & 70 & 3 & 490K \\
\qute & 184 & 111 & 73 & 1 & 515K \\
\hline
\end{tabular}
}
\label{fig:exp:463:instances:prepro:bloqqer:qrat:appendix}
}
\end{center}
\end{minipage}
\hfill
\begin{minipage}[b]{0.499\textwidth}
\begin{center}
\subfloat[\qrattool and \hqspre (QH).]{
{\setlength\tabcolsep{0.15cm}
\begin{tabular}{l@{\quad}c@{\quad}c@{\quad}r@{\quad}r@{\quad}c}
\hline
\emph{Solver} & \emph{S} & \emph{$\bot$} & \multicolumn{1}{l}{\emph{$\top$}} & \emph{U} & \emph{Time} \\
\hline
\caqefixed & 308 & 183 & 125 & 16 & 303K \\
\rareqs & 283 & 178 & 105 & 7 & 340K \\
\depqbf & 251 & 149 & 102 & 23 & 398K \\
\qute & 244 & 159 & 85 & 3 & 410K \\
\ijtihad & 231 & 159 & 72 & 2 & 428K \\
\hline
\end{tabular}
}
\label{fig:exp:463:instances:prepro:qrat:hqspre:appendix}
}
\end{center}
\end{minipage}
\label{fig:exp:463:instances:appendix}
\refstepcounter{table}
\end{table}

\end{appendix}


\end{document}